

%
%
%
\def\unredoffs{} \def\redoffs{\voffset=-.31truein\hoffset=-.59truein}
\def\speclscape{\special{ps: landscape}}
%
%
%
%
\newbox\leftpage \newdimen\fullhsize \newdimen\hstitle \newdimen\hsbody
\tolerance=1000\hfuzz=2pt
\catcode`\@=11 
\def\bigans{b }
\def\answ{b }

%
\ifx\answ\bigans\message{(This will come out unreduced.}
\magnification=1200\unredoffs\baselineskip=16pt plus 2pt minus 1pt
\hsbody=\hsize \hstitle=\hsize 
\else\message{(This will be reduced.} \let\l@r=L
\magnification=1000\baselineskip=16pt plus 2pt minus 1pt \vsize=7truein
\redoffs \hstitle=8truein\hsbody=4.75truein\fullhsize=10truein\hsize=\hsbody
\output={\ifnum\pageno=0 
  \shipout\vbox{\speclscape{\hsize\fullhsize\makeheadline}
    \hbox to \fullhsize{\hfill\pagebody\hfill}}\advancepageno
  \else
  \almostshipout{\leftline{\vbox{\pagebody\makefootline}}}\advancepageno
  \fi}
\def\almostshipout#1{\if L\l@r \count1=1 \message{[\the\count0.\the\count1]}
      \global\setbox\leftpage=#1 \global\let\l@r=R
 \else \count1=2
  \shipout\vbox{\speclscape{\hsize\fullhsize\makeheadline}
      \hbox to\fullhsize{\box\leftpage\hfil#1}}  \global\let\l@r=L\fi}
\fi
%
\newcount\yearltd\yearltd=\year\advance\yearltd by -1900

\def\Title#1#2{\nopagenumbers\abstractfont\hsize=\hstitle\rightline{#1}%
\vskip 1in\centerline{\titlefont #2}\abstractfont\vskip .5in\pageno=0}
\def\Date#1{\vfill\leftline{#1}\tenpoint\supereject\global\hsize=\hsbody%
\footline={\hss\tenrm\folio\hss}}
%

\def\draftmode{\message{ DRAFTMODE }\def\draftdate{{\rm preliminary draft:
\number\month/\number\day/\number\yearltd\ \ \hourmin}}%
\headline={\hfil\draftdate}\writelabels\baselineskip=20pt plus 2pt minus 2pt
 {\count255=\time\divide\count255 by 60 \xdef\hourmin{\number\count255}
  \multiply\count255 by-60\advance\count255 by\time
  \xdef\hourmin{\hourmin:\ifnum\count255<10 0\fi\the\count255}}}
\def\nolabels{\def\wrlabeL##1{}\def\eqlabeL##1{}\def\reflabeL##1{}}
\def\writelabels{\def\wrlabeL##1{\leavevmode\vadjust{\rlap{\smash%
{\line{{\escapechar=` \hfill\rlap{\sevenrm\hskip.03in\string##1}}}}}}}%
\def\eqlabeL##1{{\escapechar-1\rlap{\sevenrm\hskip.05in\string##1}}}%
\def\reflabeL##1{\noexpand\llap{\noexpand\sevenrm\string\string\string##1}}}
\nolabels
%
\global\newcount\secno \global\secno=0
\global\newcount\meqno \global\meqno=1
\def\newsec#1{\global\advance\secno by1\message{(\the\secno. #1)}
\global\subsecno=0\eqnres@t\noindent{\bf\the\secno. #1}
\writetoca{{\secsym} {#1}}\par\nobreak\medskip\nobreak}
\def\eqnres@t{\xdef\secsym{\the\secno.}\global\meqno=1\bigbreak\bigskip}
\def\sequentialequations{\def\eqnres@t{\bigbreak}}\xdef\secsym{}
\global\newcount\subsecno \global\subsecno=0
\def\subsec#1{\global\advance\subsecno by1\message{(\secsym\the\subsecno. #1)}
\ifnum\lastpenalty>9000\else\bigbreak\fi
\noindent{\it\secsym\the\subsecno. #1}\writetoca{\string\quad
{\secsym\the\subsecno.} {#1}}\par\nobreak\medskip\nobreak}
\def\appendix#1#2{\global\meqno=1\global\subsecno=0\xdef\secsym{\hbox{#1.}}
\bigbreak\bigskip\noindent{\bf Appendix #1. #2}\message{(#1. #2)}
\writetoca{Appendix {#1.} {#2}}\par\nobreak\medskip\nobreak}
%
%
\def\eqnn#1{\xdef #1{(\secsym\the\meqno)}\writedef{#1\leftbracket#1}%
\global\advance\meqno by1\wrlabeL#1}
\def\eqna#1{\xdef #1##1{\hbox{$(\secsym\the\meqno##1)$}}
\writedef{#1\numbersign1\leftbracket#1{\numbersign1}}%
\global\advance\meqno by1\wrlabeL{#1$\{\}$}}
\def\eqn#1#2{\xdef #1{(\secsym\the\meqno)}\writedef{#1\leftbracket#1}%
\global\advance\meqno by1$$#2\eqno#1\eqlabeL#1$$}
%
\newskip\footskip\footskip14pt plus 1pt minus 1pt 
\def\footnotefont{\ninepoint}\def\f@t#1{\footnotefont #1\@foot}
\def\f@@t{\baselineskip\footskip\bgroup\footnotefont\aftergroup\@foot\let\next}
\setbox\strutbox=\hbox{\vrule height9.5pt depth4.5pt width0pt}
\global\newcount\ftno \global\ftno=0
\def\foot{\global\advance\ftno by1\footnote{$^{\the\ftno}$}}
%
\newwrite\ftfile
\def\footend{\def\foot{\global\advance\ftno by1\chardef\wfile=\ftfile
$^{\the\ftno}$\ifnum\ftno=1\immediate\openout\ftfile=foots.tmp\fi%
\immediate\write\ftfile{\noexpand\smallskip%
\noexpand\item{f\the\ftno:\ }\pctsign}\findarg}%
\def\footatend{\vfill\eject\immediate\closeout\ftfile{\parindent=20pt
\centerline{\bf Footnotes}\nobreak\bigskip\input foots.tmp }}}
\def\footatend{}
%
%
\global\newcount\refno \global\refno=1
\newwrite\rfile
\def\ref{[\the\refno]\nref}
\def\nref#1{\xdef#1{[\the\refno]}\writedef{#1\leftbracket#1}%
\ifnum\refno=1\immediate\openout\rfile=refs.tmp\fi
\global\advance\refno by1\chardef\wfile=\rfile\immediate
\write\rfile{\noexpand\item{#1\ }\reflabeL{#1\hskip.31in}\pctsign}\findarg}
\def\findarg#1#{\begingroup\obeylines\newlinechar=`\^^M\pass@rg}
{\obeylines\gdef\pass@rg#1{\writ@line\relax #1^^M\hbox{}^^M}%
\gdef\writ@line#1^^M{\expandafter\toks0\expandafter{\striprel@x #1}%
\edef\next{\the\toks0}\ifx\next\em@rk\let\next=\endgroup\else\ifx\next\empty%
\else\immediate\write\wfile{\the\toks0}\fi\let\next=\writ@line\fi\next\relax}}
\def\striprel@x#1{} \def\em@rk{\hbox{}}
\def\lref{\begingroup\obeylines\lr@f}
\def\lr@f#1#2{\gdef#1{\ref#1{#2}}\endgroup\unskip}

\def\addref#1{\immediate\write\rfile{\noexpand\item{}#1}} 
\def\startrefs#1{\immediate\openout\rfile=refs.tmp\refno=#1}
\def\xref{\expandafter\xr@f}\def\xr@f[#1]{#1}
\def\refs#1{\count255=1[\r@fs #1{\hbox{}}]}
\def\r@fs#1{\ifx\und@fined#1\message{reflabel \string#1 is undefined.}%
\nref#1{need to supply reference \string#1.}\fi%
\vphantom{\hphantom{#1}}\edef\next{#1}\ifx\next\em@rk\def\next{}%
\else\ifx\next#1\ifodd\count255\relax\xref#1\count255=0\fi%
\else#1\count255=1\fi\let\next=\r@fs\fi\next}
%

%
\newwrite\ffile\global\newcount\figno \global\figno=1
\def\fig{fig.~\the\figno\nfig}
\def\nfig#1{\xdef#1{fig.~\the\figno}%
\writedef{#1\leftbracket fig.\noexpand~\the\figno}%
\ifnum\figno=1\immediate\openout\ffile=figs.tmp\fi\chardef\wfile=\ffile%
\immediate\write\ffile{\noexpand\medskip\noexpand\item{Fig.\ \the\figno. }
\reflabeL{#1\hskip.55in}\pctsign}\global\advance\figno by1\findarg}
\def\xfig{\expandafter\xf@g}\def\xf@g fig.\penalty\@M\ {}
\def\figs#1{figs.~\f@gs #1{\hbox{}}}
\def\f@gs#1{\edef\next{#1}\ifx\next\em@rk\def\next{}\else
\ifx\next#1\xfig #1\else#1\fi\let\next=\f@gs\fi\next}
\newwrite\lfile
{\escapechar-1\xdef\pctsign{\string\%}\xdef\leftbracket{\string\{}
\xdef\rightbracket{\string\}}\xdef\numbersign{\string\#}}

\def\writestop{\def\writestoppt{\immediate\write\lfile{\string\pageno%
\the\pageno\string\startrefs\leftbracket\the\refno\rightbracket%
\string\def\string\secsym\leftbracket\secsym\rightbracket%
\string\secno\the\secno\string\meqno\the\meqno}\immediate\closeout\lfile}}
\def\writestoppt{}\def\writedef#1{}
\def\seclab#1{\xdef #1{\the\secno}\writedef{#1\leftbracket#1}\wrlabeL{#1=#1}}
\def\subseclab#1{\xdef #1{\secsym\the\subsecno}%
\writedef{#1\leftbracket#1}\wrlabeL{#1=#1}}
\newwrite\tfile \def\writetoca#1{}
\def\leaderfill{\leaders\hbox to 1em{\hss.\hss}\hfill}
\def\writetoc{\immediate\openout\tfile=toc.tmp
   \def\writetoca##1{{\edef\next{\write\tfile{\noindent ##1
   \string\leaderfill {\noexpand\number\pageno} \par}}\next}}}

%
\catcode`\@=12 
%
\edef\tfontsize{\ifx\answ\bigans scaled\magstep3\else scaled\magstep4\fi}
\font\titlerm=cmr10 \tfontsize \font\titlerms=cmr7 \tfontsize
\font\titlermss=cmr5 \tfontsize \font\titlei=cmmi10 \tfontsize
\font\titleis=cmmi7 \tfontsize \font\titleiss=cmmi5 \tfontsize
\font\titlesy=cmsy10 \tfontsize \font\titlesys=cmsy7 \tfontsize
\font\titlesyss=cmsy5 \tfontsize \font\titleit=cmti10 \tfontsize
\skewchar\titlei='177 \skewchar\titleis='177 \skewchar\titleiss='177
\skewchar\titlesy='60 \skewchar\titlesys='60 \skewchar\titlesyss='60
\def\titlefont{\def\rm{\fam0\titlerm}
\textfont0=\titlerm \scriptfont0=\titlerms \scriptscriptfont0=\titlermss
\textfont1=\titlei \scriptfont1=\titleis \scriptscriptfont1=\titleiss
\textfont2=\titlesy \scriptfont2=\titlesys \scriptscriptfont2=\titlesyss
\textfont\itfam=\titleit \def\it{\fam\itfam\titleit}\rm}
 \ifx\answ\bigans\else scaled\magstep1\fi
\ifx\answ\bigans\def\abstractfont{\tenpoint}\else
\font\abssl=cmsl10 scaled \magstep1
\font\absrm=cmr10 scaled\magstep1 \font\absrms=cmr7 scaled\magstep1
\font\absrmss=cmr5 scaled\magstep1 \font\absi=cmmi10 scaled\magstep1
\font\absis=cmmi7 scaled\magstep1 \font\absiss=cmmi5 scaled\magstep1
\font\abssy=cmsy10 scaled\magstep1 \font\abssys=cmsy7 scaled\magstep1
\font\abssyss=cmsy5 scaled\magstep1 \font\absbf=cmbx10 scaled\magstep1
\skewchar\absi='177 \skewchar\absis='177 \skewchar\absiss='177
\skewchar\abssy='60 \skewchar\abssys='60 \skewchar\abssyss='60
\def\abstractfont{\def\rm{\fam0\absrm}
\textfont0=\absrm \scriptfont0=\absrms \scriptscriptfont0=\absrmss
\textfont1=\absi \scriptfont1=\absis \scriptscriptfont1=\absiss
\textfont2=\abssy \scriptfont2=\abssys \scriptscriptfont2=\abssyss
\textfont\itfam=\bigit \def\it{\fam\itfam\bigit}\def\footnotefont{\tenpoint}%
\textfont\slfam=\abssl \def\sl{\fam\slfam\abssl}%
\textfont\bffam=\absbf \def\bf{\fam\bffam\absbf}\rm}\fi
\def\tenpoint{\def\rm{\fam0\tenrm}
\textfont0=\tenrm \scriptfont0=\sevenrm \scriptscriptfont0=\fiverm
\textfont1=\teni  \scriptfont1=\seveni  \scriptscriptfont1=\fivei
\textfont2=\tensy \scriptfont2=\sevensy \scriptscriptfont2=\fivesy
\textfont\itfam=\tenit \def\it{\fam\itfam\tenit}\def\footnotefont{\ninepoint}%
\textfont\bffam=\tenbf \def\bf{\fam\bffam\tenbf}\def\sl{\fam\slfam\tensl}\rm}
\font\ninerm=cmr9 \font\sixrm=cmr6 \font\ninei=cmmi9 \font\sixi=cmmi6
\font\ninesy=cmsy9 \font\sixsy=cmsy6 \font\ninebf=cmbx9
\font\nineit=cmti9 \font\ninesl=cmsl9 \skewchar\ninei='177
\skewchar\sixi='177 \skewchar\ninesy='60 \skewchar\sixsy='60
\def\ninepoint{\def\rm{\fam0\ninerm}
\textfont0=\ninerm \scriptfont0=\sixrm \scriptscriptfont0=\fiverm
\textfont1=\ninei \scriptfont1=\sixi \scriptscriptfont1=\fivei
\textfont2=\ninesy \scriptfont2=\sixsy \scriptscriptfont2=\fivesy
\textfont\itfam=\ninei \def\it{\fam\itfam\nineit}\def\sl{\fam\slfam\ninesl}%
\textfont\bffam=\ninebf \def\bf{\fam\bffam\ninebf}\rm}
%
%

\hyphenation{anom-aly anom-alies coun-ter-term coun-ter-terms}
\def\inv{^{\raise.15ex\hbox{${\scriptscriptstyle -}$}\kern-.05em 1}}

\def\Dsl{\,\raise.15ex\hbox{/}\mkern-13.5mu D} 
\def\dsl{\raise.15ex\hbox{/}\kern-.57em\partial}

\font\bigit=cmti10 scaled \magstep1
\def\lspace{\ifx\answ\bigans{}\else\qquad\fi}
\def\lbspace{\ifx\answ\bigans{}\else\hskip-.2in\fi} 
\def\boxeqn#1{\vcenter{\vbox{\hrule\hbox{\vrule\kern3pt\vbox{\kern3pt
	\hbox{${\displaystyle #1}$}\kern3pt}\kern3pt\vrule}\hrule}}}
\def\mbox#1#2{\vcenter{\hrule \hbox{\vrule height#2in
		\kern#1in \vrule} \hrule}}  
%

\def\darr#1{\raise1.5ex\hbox{$\leftrightarrow$}\mkern-16.5mu #1}

\def\roughly#1{\raise.3ex\hbox{$#1$\kern-.75em\lower1ex\hbox{$\sim$}}}

\let\includefigures=\iftrue
\let\useblackboard=\iftrue
\newfam\black

\includefigures
\message{If you do not have epsf.tex (to include figures),}
\message{change the option at the top of the tex file.}
\input epsf
\def\figin{\epsfcheck\figin}\def\figins{\epsfcheck\figins}
\def\epsfcheck{\ifx\epsfbox\UnDeFiNeD
\message{(NO epsf.tex, FIGURES WILL BE IGNORED)}
\gdef\figin##1{\vskip2in}\gdef\figins##1{\hskip.5in}
\else\message{(FIGURES WILL BE INCLUDED)}%
\gdef\figin##1{##1}\gdef\figins##1{##1}\fi}
\def\DefWarn#1{}
\def\figinsert{\goodbreak\midinsert}
\def\ifig#1#2#3{\DefWarn#1\xdef#1{fig.~\the\figno}
\writedef{#1\leftbracket fig.\noexpand~\the\figno}%
\figinsert\figin{\centerline{#3}}\medskip\centerline{\vbox{
\baselineskip12pt\advance\hsize by -1truein
\noindent\footnotefont{\bf Fig.~\the\figno:} #2}}
\endinsert\global\advance\figno by1}
\else
\def\ifig#1#2#3{\xdef#1{fig.~\the\figno}
\writedef{#1\leftbracket fig.\noexpand~\the\figno}%
\global\advance\figno by1} \fi

\def\id{{1 \kern-.28em {\rm l}}}
\def\N{{\cal N}}

\def\A{{\cal A}}

\def\K3{{\bf K3}}
\def\journal#1&#2(#3){\unskip, \sl #1\ \bf #2 \rm(19#3) }
\def\andjournal#1&#2(#3){\sl #1~\bf #2 \rm (19#3) }

\def\bar{\overline}

\def\tilde{\widetilde}

\def\frac#1#2{{#1\over#2}}

\def\inbar{\,\vrule height1.5ex width.4pt depth0pt}
\def\IC{\relax\hbox{$\inbar\kern-.3em{\rm C}$}}
\def\IR{\relax{\rm I\kern-.18em R}}
\def\IP{\relax{\rm I\kern-.18em P}}

%
%

%
\catcode`\@=11
\def\slash#1{\mathord{\mathpalette\c@ncel{#1}}}
\overfullrule=0pt
\def\AA{{\cal A}}

\def\FF{{\cal F}}
\def\GG{{\cal G}}

\def\LL{{\cal L}}
\def\NN{{\cal N}}
\def\OO{{\cal O}}

\def\SS{{\cal S}}

\def\underrel#1\over#2{\mathrel{\mathop{\kern\z@#1}\limits_{#2}}}

\catcode`\@=12


%

\def\det{{\rm det}}

\def\det{{\rm det}}


\def\p{{\partial}}

\def\ra{{\rightarrow}}

\def\tmu{{\tilde \mu}}

\def\Af{{\widetilde A}}
\def\tm{{\tilde{m}}}
\def\tmu{{\tilde{\mu}}}
\def\tq{{\tilde{q}}}
\def\tom{{\tilde{\omega}}}

\def\tq{{\tilde q}}
\def\A{{\cal  A}}
\def\B{{\cal   B}}

\def\dfde{$D4-D8-\overline{D8}$}

\lref\ParnachevEV{
  A.~Parnachev and D.~A.~Sahakyan,
  ``Photoemission with chemical potential from QCD gravity dual,''
  Nucl.\ Phys.\  B {\bf 768}, 177 (2007)
  [arXiv:hep-th/0610247].
}

\lref\HorigomeXU{
  N.~Horigome and Y.~Tanii,
  ``Holographic chiral phase transition with chemical potential,''
  JHEP {\bf 0701}, 072 (2007)
  [arXiv:hep-th/0608198].
}

\lref\SonXC{
  D.~T.~Son and M.~A.~Stephanov,
  ``QCD at finite isospin density,''
  Phys.\ Rev.\ Lett.\  {\bf 86}, 592 (2001)
  [arXiv:hep-ph/0005225].
}

\lref\SakaiCN{
  T.~Sakai and S.~Sugimoto,
  ``Low energy hadron physics in holographic QCD,''
  Prog.\ Theor.\ Phys.\  {\bf 113}, 843 (2005)
  [arXiv:hep-th/0412141].
}

\lref\KarchFA{
  A.~Karch, D.~T.~Son and A.~O.~Starinets,
  ``Zero Sound from Holography,''
  arXiv:0806.3796 [hep-th].
}

\lref\KarchSH{
  A.~Karch and E.~Katz,
  ``Adding flavor to AdS/CFT,''
  JHEP {\bf 0206}, 043 (2002)
  [arXiv:hep-th/0205236].
}

\lref\McLerranQJ{
  L.~McLerran and R.~D.~Pisarski,
  ``Phases of Cold, Dense Quarks at Large $N_c$,''
  Nucl.\ Phys.\  A {\bf 796}, 83 (2007)
  [arXiv:0706.2191 [hep-ph]].
}

\lref\KimXI{
  Y.~Kim, C.~H.~Lee and H.~U.~Yee,
  ``Holographic Nuclear Matter in AdS/QCD,''
  Phys.\ Rev.\  D {\bf 77}, 085030 (2008)
  [arXiv:0707.2637 [hep-ph]].
}

\lref\BergmanWP{
  O.~Bergman, G.~Lifschytz and M.~Lippert,
  ``Holographic Nuclear Physics,''
  JHEP {\bf 0711}, 056 (2007)
  [arXiv:0708.0326 [hep-th]].
}

\lref\DavisKA{
  J.~L.~Davis, M.~Gutperle, P.~Kraus and I.~Sachs,
  ``Stringy NJL and Gross-Neveu models at finite density and temperature,''
  JHEP {\bf 0710}, 049 (2007)
  [arXiv:0708.0589 [hep-th]].
}
\lref\RozaliRX{
  M.~Rozali, H.~H.~Shieh, M.~Van Raamsdonk and J.~Wu,
  ``Cold Nuclear Matter In Holographic QCD,''
  JHEP {\bf 0801}, 053 (2008)
  [arXiv:0708.1322 [hep-th]].
}

\lref\KimZM{
  K.~Y.~Kim, S.~J.~Sin and I.~Zahed,
  ``The Chiral Model of Sakai-Sugimoto at Finite Baryon Density,''
  JHEP {\bf 0801}, 002 (2008)
  [arXiv:0708.1469 [hep-th]].
}

\lref\KimVD{
  K.~Y.~Kim, S.~J.~Sin and I.~Zahed,
  ``Dense Holographic QCD in the Wigner-Seitz Approximation,''
  JHEP {\bf 0809}, 001 (2008)
  [arXiv:0712.1582 [hep-th]].
}

\lref\BergmanQV{
  O.~Bergman, G.~Lifschytz and M.~Lippert,
  ``Magnetic properties of dense holographic QCD,''
  arXiv:0806.0366 [hep-th].
}

\lref\SakaiYT{
  T.~Sakai and S.~Sugimoto,
  ``More on a holographic dual of QCD,''
  Prog.\ Theor.\ Phys.\  {\bf 114}, 1083 (2005)
  [arXiv:hep-th/0507073].
}

\lref\KarchFA{
  A.~Karch, D.~T.~Son and A.~O.~Starinets,
  ``Zero Sound from Holography,''
  arXiv:0806.3796 [hep-th].
}

\lref\KutasovAH{
  E.~Antonyan, J.~A.~Harvey, S.~Jensen and D.~Kutasov,
  ``NJL and QCD from string theory,''
  arXiv:hep-th/0604017 [hep-th].
}

\lref\ParnachevZS{
  M.~Kulaxizi and A.~Parnachev,
  ``Comments on Fermi Liquid from Holography,''
  Phys.\ Rev.\  D {\bf 78}, 086004 (2008)
  [arXiv:0808.3953 [hep-th]].
}

\lref\LL{
  E.M. Lifshitz and L.P. Pitaevskii, Statistical Physics Part 2, Pergamon Press, Oxford, 1980.
}

\lref\PN{P. Nozieres and D. Pines, The theory of Quantum Liquids, Part I, Benjamin, NY,1966.
}

\lref\SkenderisDF{
  I.~Kanitscheider, K.~Skenderis and M.~Taylor,
  ``Precision holography for non-conformal branes,''
  JHEP {\bf 0809}, 094 (2008)
  [arXiv:0807.3324 [hep-th]].
}

\lref\KarchBR{
  A.~Karch and A.~O'Bannon,
  ``Holographic Thermodynamics at Finite Baryon Density: Some Exact Results,''
  arXiv:0709.0570 [hep-th].
}

\lref\SkenderisSVR{
   K.~Skenderis and B.~C.~van Rees,
   ``Real-time gauge/gravity duality,''
   [arXiv:hep-th/0805.0150].
}

\lref\SonSS{
    D.~T.~Son and A.~O.~Starinets,
  ``Minkowski-space correlators in AdS/CFT correspondence: Recipe and applications,''
    JHEP {\bf 0209}, 042 (2002)
   [arXiv:hep-th/0205051].
}

\lref\HerzogHS{
C.~P.~Herzog and D.~T.~Son,
  ``Schwinger-Keldysh propagators from AdS/CFT correspondence,''
   JHEP {\bf 0303}, 046 (2003)
   [arXiv:hep-th/0212072].
}

\lref\Marolf{
D.~Marolf,
  ``States and boundary terms: Subtleties of Lorentzian AdS/CFT,''
  JHEP {\bf 0505}, 042 (2005)
  [arXiv:hep-th/0412032].
}

\lref\ParnachevDN{
  A.~Parnachev and D.~A.~Sahakyan,
  ``Chiral phase transition from string theory,''
  Phys.\ Rev.\ Lett.\  {\bf 97}, 111601 (2006)
  [arXiv:hep-th/0604173].
}

\lref\SkenderisHR{
  K.~Skenderis,
  ``Lecture notes on holographic renormalization,''
  Class.\ Quant.\ Grav.\  {\bf 19}, 5849 (2002)
  [arXiv:hep-th/0209067].
}

\lref\MasNG{
  J.~Mas and J.~Tarrio,
  ``Hydrodynamics from the Dp-brane,''
  JHEP {\bf 0705}, 036 (2007)
  [arXiv:hep-th/0703093].
}

\lref\CaceresTA{
  E.~Caceres, M.~Natsuume and T.~Okamura,
  ``Screening length in plasma winds,''
  JHEP {\bf 0610}, 011 (2006)
  [arXiv:hep-th/0607233].
}

\lref\MyersTB{
  D.~Mateos, R.~C.~Myers and R.~M.~Thomson,
  ``Thermodynamics of the brane,''
  JHEP {\bf 0705}, 067 (2007)
  [arXiv:hep-th/0701132].
}

\lref\SongRFL{
  C.~Song,
  ``Dense nuclear matter: Landau Fermi-liquid theory and chiral Lagrangian
  with scaling,''
  Phys.\ Rept.\  {\bf 347}, 289 (2001)
  [arXiv:nucl-th/0006030].
}

\lref\BaymRFL{
  G.~Baym and S.~A.~Chin,
  ``Landau Theory Of Relativistic Fermi Liquids,''
  Nucl.\ Phys.\  A {\bf 262}, 527 (1976).
}

\lref\AharonyAHA{
  O.~Aharony, J.~Sonnenschein and S.~Yankielowicz,
  ``A holographic model of deconfinement and chiral symmetry restoration,''
  Annals Phys.\  {\bf 322}, 1420 (2007)
  [arXiv:hep-th/0604161].
}

\lref\KimBV{
  K.~Y.~Kim and I.~Zahed,
  ``Baryonic Response of Dense Holographic QCD,''
  arXiv:0811.0184 [hep-th].
}

\lref\HataMB{
  H.~Hata, T.~Sakai, S.~Sugimoto and S.~Yamato,
  ``Baryons from instantons in holographic QCD,''
  arXiv:hep-th/0701280.
}

\lref\DavisNV{
  J.~L.~Davis, P.~Kraus and A.~Shah,
  ``Gravity Dual of a Quantum Hall Plateau Transition,''
  arXiv:0809.1876 [hep-th].
}

\lref\LeeXF{
  S.~S.~Lee,
  ``A Non-Fermi Liquid from a Charged Black Hole: A Critical Fermi Ball,''
  arXiv:0809.3402 [hep-th].
}

\lref\MyersME{
  R.~C.~Myers and M.~C.~Wapler,
  ``Transport Properties of Holographic Defects,''
  arXiv:0811.0480 [hep-th].
}

\lref\ParnachevBC{
  A.~Parnachev,
  ``Holographic QCD with Isospin Chemical Potential,''
  JHEP {\bf 0802}, 062 (2008)
  [arXiv:0708.3170 [hep-th]].
}

\lref\NambuTP{
  Y.~Nambu and G.~Jona-Lasinio,
  ``Dynamical model of elementary particles based on an analogy with
  superconductivity. I,''
  Phys.\ Rev.\  {\bf 122}, 345 (1961).
}

\lref\OBannonSC{
  A.~O'Bannon,
  ``Toward a Holographic Model of Superconducting Fermions,''
  arXiv:0811.0198 [hep-th].
}

\lref\GubserSC{
  S.~S.~Gubser,
  ``Breaking an Abelian gauge symmetry near a black hole horizon,''
  arXiv:0801.2977 [hep-th].
}

\lref\GubserSP{
  S.~S.~Gubser and S.~S.~Pufu,
  ``The gravity dual of a p-wave superconductor,''
  arXiv:0805.2960 [hep-th].
}

\lref\GubserAN{
  S.~S.~Gubser and A.~Nellore,
  ``Low-temperature behavior of the Abelian Higgs model in anti-de Sitter
  space,''
  arXiv:0810.4554 [hep-th].
}

\lref\HartnollHH{
  S.~A.~Hartnoll, C.~P.~Herzog and G.~T.~Horowitz,
  ``Building a Holographic Superconductor,''
  Phys.\ Rev.\ Lett.\  {\bf 101}, 031601 (2008)
  [arXiv:0803.3295 [hep-th]].
}

\lref\RobertsSH{
  M.~M.~Roberts and S.~A.~Hartnoll,
  ``Pseudogap and time reversal breaking in a holographic superconductor,''
  JHEP {\bf 0808}, 035 (2008)
  [arXiv:0805.3898 [hep-th]].
}

\lref\HartnollSHGH{
  S.~A.~Hartnoll, C.~P.~Herzog and G.~T.~Horowitz,
  ``Holographic Superconductors,''
  arXiv:0810.1563 [hep-th].
}

\lref\MaedaTO{
  K.~Maeda and T.~Okamura,
  ``Characteristic length of an AdS/CFT superconductor,''
  arXiv:0809.3079 [hep-th].
}

\lref\BasuHPS{
  P.~Basu, J.~He, A.~Mukherjee and H.~H.~Shieh,
  ``Superconductivity from D3/D7: Holographic Pion Superfluid,''
  arXiv:0810.3970 [hep-th].
}

\lref\AmmonSCF{
  M.~Ammon, J.~Erdmenger, M.~Kaminski and P.~Kerner,
  ``Superconductivity from gauge/gravity duality with flavor,''
  arXiv:0810.2316 [hep-th].
}

\lref\ShiehNF{
  H.~H.~Shieh and G.~van Anders,
  ``Comments on Holographic Fermi Surfaces,''
  arXiv:0810.1661 [hep-th].
}

\lref\KimGP{
  K.~Y.~Kim, S.~J.~Sin and I.~Zahed,
  ``Dense hadronic matter in holographic QCD,''
  arXiv:hep-th/0608046.
}

\lref\LeighTT{
  R.~G.~Leigh, N.~N.~Hoang and A.~C.~Petkou,
  ``Torsion and the Gravity Dual of Parity Symmetry Breaking in AdS4/CFT3
  Holography,''
  arXiv:0809.5258 [hep-th].
}

\lref\NawaGV{
  K.~Nawa, H.~Suganuma and T.~Kojo,
  ``Baryons in Holographic QCD,''
  Phys.\ Rev.\  D {\bf 75}, 086003 (2007)
  [arXiv:hep-th/0612187].
}

\lref\NawaUV{
  K.~Nawa, H.~Suganuma and T.~Kojo,
  ``Brane-induced Skyrmion on $S^3$: baryonic matter in holographic QCD,''
  arXiv:0810.1005 [hep-th].
}

\Title{\vbox{\baselineskip12pt
}}
{\vbox{\centerline{Holographic Responses of Fermion Matter}
\vskip.06in
}}
\centerline{Manuela Kulaxizi${}^a$ and  Andrei Parnachev${}^b$}
\bigskip
\centerline{{\it ${}^a$Institute for Theoretical Physics, University of Amsterdam,}}
\centerline{{\it Valckenierstraat 65, 1018XE Amsterdam, The Netherlands }}
\centerline{{\it ${}^b$C.N.Yang Institute for Theoretical Physics, Department of Physics,}}
\centerline{{\it Stony Brook University, Stony Brook, NY 11794-3840, USA}}
\vskip.1in \vskip.1in \centerline{\bf Abstract}
\noindent
We consider the  $D4-D8-\overline{D8}$ brane system which serves as ultraviolet completion
of the Nambu-Jona-Lasinio model, where the only degrees of freedom carrying
baryon charge are fermions.
By turning on chemical potential for this charge one  may expect
the formation of the Fermi liquid ground state.
At strong coupling we use the dual
holographic description to investigate the responses of the system to small perturbations.
In the chirally symmetric phase we
find that the density dependent part of the heat capacity
vanishes linearly with temperature.
We also observe a zero sound excitation in the collisionless regime,
whose speed is equal to that of normal sound in the hydrodynamic regime.
Both the linear dependence of the heat capacity and the existence of zero
sound are properties of the Fermi liquid ground state.
We also compute the two-point function of the currents at vanishing frequency
but do not find any singularities at finite values of the momentum.

\vfill

\Date{December 2008}


\newsec{Introduction}\seclab\sintro

Understanding QCD at finite baryon density or  finite chemical
potential is an important and difficult problem.
Lattice computations are complicated by the sign problem, and one largely
has to rely on phenomenological models.
At large values of the chemical potential asymptotic freedom ensures
that the physics is determined by the dynamics of the quarks near the Fermi
surface.
The physics simplifies in the planar limit, since
some of the perturbative instabilities are suppressed
(see \McLerranQJ\ and references therein  for the recent work
in this direction.).

It is interesting to compare the predictions of perturbative QCD with
those coming from string theoretic models.
Consider the holographic model of QCD \refs{\SakaiCN,\SakaiYT} at finite temperature $T$
and baryon chemical potential $\mu$.
An incomplete list of references where this setup was analyzed includes
\refs{\KimGP\HorigomeXU\ParnachevEV\NawaGV\HataMB\KimXI\BergmanWP\DavisKA\RozaliRX\KimZM\KimVD-\NawaUV}.
The holographic model is by no means equivalent to QCD, and in fact
lacks asymptotic freedom.
In the holographic model it is possible
to adjust the glueball mass scale to be lower than the meson mass scale.
This leads to the
existence of the deconfined but chirally broken phase.
The phase diagram then contains a line of first order phase transitions
between a chirally symmetric phase at larger values of $\mu$ and $T$
and a broken phase at smaller values of these parameters \refs{\HorigomeXU,\ParnachevEV}.
In addition, at large values of $\mu$ there is a phase with condensed
baryons \refs{\BergmanWP,\RozaliRX}, which is not what we naively expect from perturbative QCD.
One may ask whether the physics of the system resembles that of
perturbative QCD, and, in particular whether one can see signatures of
Fermi liquid formation.

The question is actually more general that this\foot{Recent work on holographic
Fermi systems includes \refs{\ParnachevBC\DavisNV\LeeXF\ShiehNF-\MyersME}, see
also \LeighTT.}. Generic attractive
interactions are believed to destabilize the Fermi surface, leading to the
formation of a gap and a superconducting ground state. (For a holographic description
of superconductivity see \refs{\GubserSC\GubserSP\GubserAN\HartnollHH\RobertsSH\HartnollSHGH\MaedaTO\AmmonSCF\BasuHPS-\OBannonSC} 
and references therein.)
In the 't Hooft limit the perturbative quark-quark interactions
are suppressed, and there is an opportunity for the Fermi liquid ground
state to survive, even though the coupling is strong.
In this paper we attempt to analyze the fate of
the fermion matter from the $D4-D8$ strings
by studying its responses to small external perturbations.
Since our main interest is not QCD dynamics and the Yang-Mills
degrees of freedom simply provide the strong interaction between quarks,
we consider the (decompactification) limit which
corresponds, in the field theoretic regime, to a certain UV
completion of the Nambu-Jona-Lasinio model \NambuTP.
(See \KutasovAH\ for more details.)
We turn on chemical potential which is expected to lead to the
formation of charged matter, whose ingredients, at least in
the field theoretic regime are fundamental fermions.
We consider  the phase where chiral symmetry is restored, which
corresponds to the $D8-\overline{D8}$ branes falling into the horizon
of the black hole.
Such a phase is perturbatively stable for a wide range of
$\mu$ and $T$.

We show that the density dependent part of the heat capacity
at low temperature is linear in $T$.
This is the behavior expected for systems with a Fermi surface,
where only a fraction of quasiparticles is excited at small temperatures.
We also observe that at arbitrarily low temperatures there exists a massless excitation
(zero sound) whose speed is equal to the speed of normal sound in the
hydrodynamic regime.
The existence of zero sound is also a feature of the Fermi liquid and corresponds to
the deformation of the Fermi surface away from the spherical shape.
We then compute the  current-current two-point function at vanishing
frequency $\omega=0$ and finite spacial momentum $q$
using the holographic techniques.
Such a two-point function should be sensitive to the finite gap
in the distribution function at the Fermi momentum $q=q_F$.
We do not find any singular features however.
As we discuss below, this might be related to a non-generic dispersion relation
of the quasiparticles near the Fermi surface.

The rest of the paper is organized as follows.
In the next section we review the $D4-D8-\overline{D8}$ system
at finite chemical potential, including phase structure and thermodynamics.
We compute the speed of normal sound in the hydrodynamic regime and
show that the density dependent part of the heat
capacity vanishes linearly with $T$ at small temperatures.
We then study small fluctuations at small temperature
(collisionless regime) and find the massless excitation (zero sound) in section 3.
The speed of zero sound is equal to the speed of normal sound in the
hydrodynamic regime.
In section 4 we compute the current-current two-point function and
show that no visible singularities are present.
We discuss our results in section 5.
Appendix A contains the derivation of the speed of normal sound in the
$D3-D7$ system.
Appendix B is devoted to the derivation of the boundary conditions in the
limit of vanishing $\omega$.

\newsec{Thermodynamics} \seclab\sthermo

\subsec{Review}\seclab\srev

In this section we will review the system studied in \refs{\KutasovAH} at zero temperature.
The configuration we are interested in consists of three kinds of branes:
$D$4, $D$8 and $\bar{D}$8 branes. The $N_c$ $D$4 branes are extended along the $x^0,\cdots,x^4$ directions
whereas the $N_f$ $D$8 and $\bar{D}$8 branes are parallel and located a distance $L$ apart from each other
along the $x^4$ direction. The only massless excitations of the fundamental strings stretching between
the  $D$8 and $D$4 branes are spacetime Weyl fermions. In particular, $4-8$ strings contribute left-handed
$q_L$ fermions localized at the $x_4=-{L\over 2}$ intersection while $4-\bar{8}$ strings provide right-handed
$q_R$ fermions localized at $x_4={L\over 2}$. Studying this system at strong coupling, i.e., $\lambda\gg L\gg l_s$
where $\lambda$ is the five-dimensional 't Hooft coupling constant, requires taking the near horizon limit of
the $D$4-branes. Assuming $N_f\ll N_c$, the back-reaction of the $D$8 branes on the geometry can be neglected.
In this case, it suffices to explore the physics of the fundamental matter through the DBI (Dirac-Born-Infeld) action
governing their dynamics as they propagate in the near horizon geometry of the $N_c$ $D$4 branes.

In the following, we summarize some of the results of \ParnachevEV\ concerning the $D4-D8-\bar{D}8$ system at zero temperature
but non-zero chemical potential. This requires turning on an electric field on the worldvolume of the $D$8
branes. The value of the chemical potential is then read off from the asymptotic value of the gauge field on the brane.

Let us start by considering the near horizon region of D4-branes
\eqn\dfournh{\eqalign{ds^2&=\left({U\over R}\right)^{{3\over 2}} \left(-dt^2+dx_i^2+dX_4^2\right)+
              \left({U\over R}\right)^{-{3\over 2}}\left(dU^2+U^2 d\Omega_4^2\right)\cr
              e^{\Phi}&=g_s \left(U\over R\right)^{{3\over 4}} \qquad\qquad\qquad F_4=dC_3={2\pi N_c\over \Omega_4}\omega_4 }}
Here $t$ is time, $i=1,2,3$ are the three spatial dimensions of the worldvolume of the D4--brane, $U$ is the radial
direction and $d\Omega_4^2$ the metric of the unit four-sphere. $\Omega_4$ and $\omega_4$ denote the
volume and volume form of the unit four-sphere respectively. The parameter $R$ in \dfournh\ is defined as
\eqn\Rdef{R^3=\pi g_s N_c l_s^3=\pi \lambda}
where $\lambda$ is the 't Hooft coupling constant. Note that here and in the rest of the paper we set $\alpha'=1$.

$D$8 branes propagating in the geometry \dfournh\ wrap $R^{3,1}\times S^4$ and their embedding profile
is specified by a single function $X_4(U)$. Assuming chiral symmetry is restored $\p_U X_4=0$ and the
induced metric takes the following form
\eqn\Deightinduced{ds_{D8}^2=\left({U\over R}\right)^{{3\over 2}} \left(-dt^2+dx_i^2\right)+
              \left({U\over R}\right)^{-{3\over 2}}\left(dU^2+U^2 d\Omega_4^2\right)}
Studying the system at finite chemical potential, requires introducing non-trivial flux $F_{0U}\neq 0$
along the brane worldvolume. The action density (DBI) reads
\eqn\DBIbg{S_{D8}=-\NN \int d U U^{{5\over 2}} \sqrt{1-(\p_U A_0)^2}}
where $\NN\equiv {\mu_8\over g_s}\Omega_4 R^{{3\over 2}}={\sqrt{2}\over 3} (2\pi)^{-{11\over 2}} {N_c N_f \over \sqrt{\lambda}}$.
Note that we made the gauge choice $A_U=0$ and rescaled $\A_0$ according to
\eqn\rescaleanot{A_0\ra 2\pi A_0}
The conserved charge associated with $A_0$ is given by
\eqn\eqmotionAnot{U^{{5\over 2}} {\p_U A_0\over \sqrt{1-(\p_U A_0)^2}}=d}
It follows that the electric field on the brane worldvolume satisfies
\eqn\soltauanot{\p_U A_0={d\over\sqrt{d^2+U^5}}}
We then determine the chemical potential $\mu$ from the asymptotic
value of $A_0$ as
\eqn\mudef{\mu={1\over 2\pi}\int_0^{\infty} \p_U A_0={1\over 2\pi}\gamma d^{{2\over 5}} }
where $\gamma$ is defined by
\eqn\gammadef{\gamma\equiv{\Gamma\left[{3\over 10}\right]\Gamma\left[{1\over 5}\right]\over 5 \sqrt{\pi}}}

\subsec{The speed of sound}\seclab\ssound

Holography relates the brane action to the grand canonical potential $\Xi=F-\mu\rho$, where $F$ the free
energy density, $\mu$ the chemical potential and $\rho$ the charge density. The precise identification at zero temperature is
$S_{D8}=-\Xi$. Evaluating the DBI action on the solution \soltauanot\ we deduce that
the grand canonical potential for this phase is given by
\eqn\renactionxmu{\Xi=-S_{D8}=-{2\over 7}\NN \gamma d^{{7\over 5}}}
where $d$ can be expressed in terms of $\mu$ through \mudef.
Note that to arrive at \renactionxmu\ we renormalized the action by subtracting the contribution
from the configuration with straight branes and no flux.

Knowledge of the grand canonical potential allows us to determine several thermodynamic quantities
of the system. The charge density for instance can be shown to be
\eqn\density{\rho=-{\delta \Xi\over\delta \mu}=2\pi d\NN}
Another interesting property is the speed of sound.
This is given by
\eqn\defsound{u^2=\left({\p P\over \p \epsilon}\right)_{\rho}}
where $P$ and $\epsilon$ denote the pressure and energy density respectively while
the derivative is taken at constant volume and particle number.
At zero temperature, the pressure is equal and opposite to the grand canonical potential $P=-\Xi$
while the energy density $\epsilon$ is equal to the free energy and is given by
\eqn\pepsilon{\epsilon=\Xi+\mu \rho={5\over 7} \NN \gamma d^{{5\over 7}}={5\over 2} P}
Using \defsound\ we find that
\eqn\soundvelocity{u^2={2\over 5}}

\subsec{Specific Heat}\subseclab\cv

The specific heat capacity $C_v$ is an important physical property of matter which often reveals the
nature of quasiparticle excitations. The specific heat of a fermionic liquid for instance, exhibits linear behavior
at low temperatures. On the other hand, one finds for a bosonic gas in $3+1$ dimensions that $C_v\sim T^3$.
In the following we will see that the heat capacity of the  $D4-D8-\bar{D}8$ system varies
linearly with the temperature in accord with the predictions of the theory of Fermi Liquids \LL.

The $D4-D8-\bar{D}8$ at finite temperature and chemical potential  has been studied in \ParnachevEV.
Here, we will briefly review the results of \ParnachevEV\ necessary for the computation of
the specific heat. The relevant background geometry is given by
\eqn\dfournhT{\eqalign{ds^2&=\left({U\over R}\right)^{{3\over 2}} \left(-f(U)dt^2+dx_i^2+dX_4^2\right)+
              \left({U\over R}\right)^{-{3\over 2}}\left({dU^2\over f(U)}+U^2 d\Omega_4^2\right)\cr
              e^{\Phi}&=g_s \left(U\over R\right)^{{3\over 4}} \qquad\qquad\qquad F_4=dC_3={2\pi N_c\over \Omega_4}\omega_4 }}
with $f(U)=1-{U_T^3\over U^3}$. The temperature $T$ is related to the minimum value $U_T$ of $U$ as
\eqn\TUT{T={3 U_T^{1\over 2}\over 4\pi R^{{3\over 2}}}\qquad\Rightarrow\qquad U_T=\left({4\pi\over 3}\right)^2 R^3 T^2}
The induced metric on the straight D8--branes is then
\eqn\DeightinducedT{ds_{D8}^2=\left({U\over R}\right)^{{3\over 2}} \left(-f(U)dt^2+dx_i^2\right)+
         \left({U\over R}\right)^{-{3\over 2}}\left({dU^2\over f(U)}+U^2 d\Omega_4^2\right)}
while the DBI action for this configuration at finite chemical potential reads
\eqn\DBIbgT{S_{D8}=-{\NN V_3\over T} \int d U U^{{5\over 2}} \sqrt{1-(\p_U A_0)^2}}
where $V_3$ denotes the volume of $R^3$. The equation of motion following from \DBIbgT\ has the exact same
form as the one at zero temperature. Eq. \soltauanot\ is therefore still valid, however the
chemical potential $\mu$ is now given by
\eqn\mudefT{\mu={1\over 2\pi}\int_{U_T}^{\infty} \p_U A_0={d\over 3\pi U_T^{{3\over 2}}}
                 F\left[{1\over 2},{3\over 10},{13\over 10},-{d^2\over U_T^5}\right]}
with $F\left[{1\over 2},{3\over 10},{13\over 10},-{d^2\over U_T^5}\right]$ the standard hypergeometric function.

Evaluating the specific heat, as well as the charge density, requires knowledge
of the grand canonical potential $\Xi$ of the system. At finite temperature, the latter
is related to the DBI action through $\Xi=-T S_{D8}$. As usual however, the action evaluated on
the solution \soltauanot\ is infinite. Holographic renormalization \SkenderisHR, \SkenderisDF\
is then required for a consistent removal of the divergences. Nonetheless, for the purposes of calculating
the charge density dependent terms of the grand canonical potential any renormalization
scheme will suffice. This is because turning on a non-trivial chemical potential
does not introduce additional divergences into the action. The appropriate counterterms are thus fixed
and independent of the charge density or chemical potential of the system\foot{This can be mainly attributed to
the flat brane embedding considered in this paper. We thank M.Taylor for explaining this to us.}.

We therefore proceed to renormalize the action by subtracting the contribution of straight branes
with no flux. The result is
\eqn\XiR{\Xi=\NN V_3 \int_{U_T}^{\infty} dU \left({U^5 \over\sqrt{U^5+d^5}}-U^{{5\over 2}}\right)=
             \NN V_3 {2\over 7} U_T^{{7\over 2}}\left[1- F\left[{1\over 2},-{7\over 10},{3\over 10},-{d^2\over U_T^5}\right]\right]}
We are interested in the density dependent term of eq. \XiR\ which we subsequently define as
\eqn\DeltaXi{\Delta\Xi= -\NN V_3 {2\over 7} U_T^{{7\over 2}}F\left[{1\over 2},-{7\over 10},{3\over 10},-{d^2\over U_T^5}\right]}
It is now straightforward to compute the charge density $\rho$ from
\eqn\rhoden{\rho\equiv - {\p \Delta\Xi\over \p \mu}=2\pi\NN d}
The specific heat on the other hand is defined through
\eqn\Cv{C_v\equiv T\left({\p S\over\p T}\right)_{\rho,V}}
where the entropy $S$ is given by
\eqn\SS{S\equiv -\left({\p \Delta\Xi\over \p T}\right)_{\mu}=-\left[\left({\p \Delta\Xi\over \p T}\right)_{d}+
              \left(\p \Delta\Xi\over \p d\right)_T \left(\p d\over \p T\right)_{\mu} \right] }
Using \mudefT\ along with the identity
\eqn\id{\left({\p d\over\p T}\right)_{\mu} \left({\p \mu\over\p d}\right)_{T} \left({\p T\over\p \mu}\right)_{d}=-1}
we find that the density-dependent part of the specific heat\foot{The density independent part behaves like $T^{6}$.}
behaves at low temperatures like
\eqn\CvdlowT{C_v\simeq \alpha T+\OO(T^{11}) \qquad
             \alpha\equiv {16\pi^2\over 9}\lambda\rho V_3  }

\newsec{Small Fluctuations and Zero Sound}\seclab\sfluct

In this section we will compute the massless excitation coupled to the
density operator in the $D4-D8-\bar{D}8$ system at strong coupling. This requires analyzing
the linearized equations of motions which follow from the action describing the dynamics of
$D$8 and $\bar{D}$8 branes.
The full action consists of a Dirac-Born-Infeld (DBI) and
a Chern-Simons (CS) term.

The DBI part of the action can be expressed as
\eqn\DBI{S_{DBI,D8}\sim\int d\Omega_4\int d^4x\int dU e^{-\Phi}\sqrt{-\det{\left[\GG+\FF\right]}}}
where $\Phi$ is the dilaton, $\GG$ the induced metric and $\FF$ the gauge field strength.
We will consider fluctuations which are independent of the coordinates of $S^4$.
Given that metric and gauge field perturbations decouple, it suffices to set $\GG=G^{(0)}$ and
expand the gauge field $\AA_0=A_0^{(0)}+A_0,\AA_i=A_i$ and the gauge field strength
$\FF_{0U}=F_{0U}^{(0)}+F_{0U},\FF_{ij}=F_{ij}$. Here $i,j=0,1,2,3$ and the superscript $(0)$ denotes the
background values of the corresponding fields. Moreover, we set $\AA_{\theta}=0, \forall~\theta\in S^4$
and choose the gauge $\AA_U=0$.

The DBI part of the action for the fluctuating fields is then given by
\eqn\DBIfl{\eqalign{S_{DBI, fl}=-{1\over 2} \N  \int d^4x \int dU g(U) \left[ \sum_{i} F_{iU}^2 - f_1(U) F_{0U}^2-f_2(U) R^3 \sum_{i} F_{0i}^2+ f_3(U)R^3\sum_{i<j} F_{ij}^2 \right]}}
where the functions $g(U)$, $f_i(U)$ are defined as
\eqn\defbg{\eqalign{g(U)&=\sqrt{U^5+d^2} \qquad f_1(U)={U^5+d^2\over U^5} \cr
f_2(U)&={1\over U^3}\qquad\qquad f_3(U)={1\over U^3} {U^5 \over U^5+d^2}}}
Not all of these functions are independent from each other. In particular, $f_2=f_1 f_3$.

We now turn to the contribution of the CS term to the action
\eqn\CSterm{S_{CS}=i{\mu_{8}\over 3!}\int_{D8} F_4\wedge \AA\wedge\FF\wedge\FF=
                  i{\mu_{8} (2\pi)^3\over 3!} N_c\int \AA\wedge\FF\wedge\FF }
where in the last equality we integrated the RR four form $F_4$ over
the four sphere. Recall that the gauge field components $\AA_I$  with $I=0,1,2,3,U$ are rescaled
according to $\AA_I\ra 2\pi \AA_I$.
Expanding \CSterm\ to quadratic order in the fields we arrive at
\eqn\CStermfl{S_{CS,fl}= i 4 d R^{{3\over 2}} {1\over 2}\NN
              \int d^4x\int dU g(U){f_3(U)\over U^2} \left[A_1 F_{23}+A_2 F_{31}+A_3 F_{12}\right] }
With the help of \CStermfl\ and \DBIfl\ we can finally write the full action for the D8 branes as
\eqn\actionfull{\eqalign{S_{D8,fl}=-{1\over 2} \N  \int d^4x \int dU g(U)
              &\left[ \sum_{i} F_{iU}^2 - f_1(U) F_{0U}^2-f_2(U) R^3 \sum_{i} F_{0i}^2+ f_3(U)R^3\sum_{i<j} F_{ij}^2- \right.\cr
&\left. - i 4 d R^{{3\over 2}} {f_3(U)\over U^2} \left(A_1 F_{23}+A_2 F_{31}+A_3 F_{12}\right)  \right] }}

It is convenient to express the equations of motion in the momentum space representation where
\eqn\mom{A_M(x^{\mu},r)=\int {d^4k\over (2\pi)^4} e^{i k_{\mu}x^{\mu}} \Af_M(k^\mu,r) }
Choosing $k_{\mu}=(-\omega,0,0,q)$ we arrive at the following set of
equations
\eqn\eqofm{\eqalign{&\p_U [g(U) f_1(U) (\p_U \Af_0)]- q^2 R^3 g(U) f_2(U) \Af_0 -q \omega R^3 g(U) f_2(U) \Af_{\parallel}=0 \cr
&\p_U [g(U) (\p_U \Af_{\parallel})]+ \omega^2 R^3 g(U) f_2(U) \Af_{\parallel}+q \omega R^3 g(U) f_2(U) \Af_0=0\cr
&\p_U [g(U) (\p_U \Af_{\perp,1})]+ g(U) R^3 [\omega^2  f_2(U)-q^2 f_3(U)] \Af_{\perp,1}-4dR^{{3\over 2}} q g(U) {f_3(U)\over U^2}\Af_{\perp,2}=0\cr
&\p_U [g(U) (\p_U \Af_{\perp,2})]+ g(U) R^3 [\omega^2  f_2(U)-q^2 f_3(U)] \Af_{\perp,2}+4dR^{{3\over 2}} q g(U) {f_3(U)\over U^2}\Af_{\perp,1}=0 }}
To fix the residual gauge invariance we additionally impose Gauss law:
\eqn\gauss{\tq\Af'_{\parallel}+\tom f_1 \Af'_{0}=0}
Henceforth we will focus on the longitudinal modes of the gauge field. Note that the CS part of the action
affected the field equations for the transverse components only.

Computing the quasinormal spectrum requires working with the gauge invariant combination
\eqn\gaugeinvarinate{E=q\Af_0+\omega \Af_{\parallel}}
We therefore use Gauss Law to express the first derivative of $\Af_0$ in terms of $E$
\eqn\eprime{\Af_0'={ q\over q^2-\omega^2 f_1} E'}
and combine the equations for the longitudinal modes in one
\eqn\eqfore{E''+\left({g'\over g}+{f'_1\over f_1} {q^2\over q^2 -\omega^2 f_1}\right)E'+R^3 f_3(\omega^2 f_1-q^2)E=0}
Primes indicate differentiation with respect to the variable $U$. Note that to arrive at \eqfore\ we also used
the identity $f_2=f_1 f_3$.

It is convenient to make a change of variables
\eqn\Utoy{y={2 \sqrt{R^3\over U}}\qquad U={4R^3\over y^2}}
to express eq. \eqfore\ as
\eqn\eqrorlong{\ddot{E}+\left({3\over y}+{\dot{g}\over g}+{\dot{f}_1\over f_1} {q^2\over q^2 -\omega^2 f_1}\right)\dot{E}+
                 {1\over f_1}(\omega^2 f_1-q^2)E=0}
Dots denote differentiation with respect to $y$ while
\eqn\functionsan{\eqalign{{\dot{g}\over g}=-{5\over y}\left({1\over 1+ \tmu^{10} y^{10}}\right) \qquad\qquad f_1=1+\tmu^{10} y^{10}\qquad\qquad
                  {\dot{f}_1\over f_1}={10\over y}\left({\tmu^{10} y^{10}\over 1+\tmu^{10} y^{10}}\right) }}
Here $\tmu$ is defined as
\eqn\tmudef{\tmu\equiv  {d^{{1\over 5}}\over 2 R^{{3\over 2}}}=\sqrt{{\mu\over 2\gamma \lambda}}}

In the vicinity of the horizon \eqrorlong\ reduces to a Bessel-type differential equation
\eqn\eqh{\ddot{E}+{3\over y}\dot{E}+\omega^2 E=0}
with general solution in terms of Hankel functions
\eqn\eqhsolgeneral{E(y)= A {H_1^{(1)}(y)\over y}+B {H_1^{(2)}(y)\over y}}
Imposing the incoming wave boundary condition at the horizon \refs{\SonSS\HerzogHS\Marolf-\SkenderisSVR} singles out one of the solutions
\eqn\eqhsol{E(y)=A {H_1^{(1)}(y)\over y}}
where $H^{(1)}$ denotes the Hankel function of first kind.
In the limit of small frequencies, or to be precise for $\omega y\ll 1$, \eqhsol\ further reduces to
\eqn\eqhsolsmallw{E(y)\simeq {A\over y^2}+A \omega^2 \left[{1\over 4}\left(1-2\tilde{\gamma}+i \pi+2 \ln{2}\right)-{1\over 2}\ln{\omega y}\right]}
with $\tilde{\gamma}$ the Euler number $\tilde{\gamma}\simeq.5772$.

On the other hand, for sufficiently small $\omega$ and $q$ the last term in eq. \eqrorlong\ can be neglected. This yields
\eqn\eqsmallw{\ddot{E}+\left({3\over y}+{\dot{g}\over g}+{\dot{f}_1\over f_1} {q^2\over q^2 -\omega^2 f_1}\right)\dot{E}=0}
Eq. \eqsmallw\ is analytically tractable. Its general solution can be exressed in terms of hypergeometric functions as follows
\eqn\solsmallw{E(y)=C_0+C_1 y^3 \left({1\over\sqrt{1+\tmu^{10} y^{10}}}-{5\over 3}\left({\omega^2\over q^2}-{2\over 5}\right)
                   F\left[{3\over 10},{1\over 2},{13\over 10},-\tmu^{10} y^{10}\right]\right)}
Near the boundary in particular \solsmallw\ reduces to
\eqn\solsmallwbnd{E(y)\simeq C_0+{5\over 3} C_1 \left(1-{\omega^2\over q^2}\right)y^3}
Imposing normalizability translates to $C_0=0$. Hence, the spectrum of quasinormal modes will be obtained
as a solution to this equation.

In the vicinity of the horizon the behavior of \solsmallw\ is
\eqn\solsmallwhor{E(y)\simeq (C_0+b C_1)+ {a C_1\over y^2}}
with $a$ and $b$ defined as follows
\eqn\abdef{a={5\over 2 \tmu^5}{\omega^2\over q^2}  \qquad b={5\over 2 \tmu^3}\gamma \left({2\over 5}-{\omega^2\over  q^2}\right)
                \qquad {b\over a}=\tmu^2 {q^2\over\omega^2}\left({2\over 5}-{\omega^2\over q^2}\right)}

Our next step would be to match the near horizon solution \solsmallwhor\ to \eqhsolsmallw.
Observe however that a logarithmic term present in \eqhsolsmallw\ is absent from \solsmallwhor.
The apparent inconsistency is resolved by computing the first order correction to the solution given by \solsmallwhor.
In the vicinity of the horizon this is a pretty simple task.
It essentially requires solving the following inhomogeneous differential equation
\eqn\difeqcorr{\ddot{E}+{3\over y}\dot{E}=-\omega^2 \left(C_0+b C_1+ {a C_1\over y^2}\right)}
The solution for large $y$ is now given by
\eqn\solsmallwhorcor{E(y)= (C_0+b C_1)+ {a C_1\over y^2}-{1\over 2} a C_1 \omega^2 \ln{y}+\OO(y^2) }
which we can readily compare with \eqhsolsmallw\ to arrive at
\eqn\matching{C_0=A\left[{1\over 4}\left(1-2\tilde{\gamma}+i \pi-2 \ln{{\omega\over 2}}\right)\omega^2-{b\over a}\right]\qquad C_1={A\over a}}

We are interested in the quasinormal mode with linear dispersion relation in the regime of small $\omega$ and $q$.
Imposing $C_0=0$ implies
\eqn\eqqn{{1\over 4}\left(1-2\tilde{\gamma}+i \pi-2 \ln{{\omega\over 2}}\right)\omega^4-{2\over 5} \tmu^2 q^2+ \tmu^2 \omega^2=0}
Neglecting higher order terms in $\omega$ and $q$ finally leads to
\eqn\res{\omega^2=u_0^2 q^2 \qquad u_0^2={2\over 5}}
This is exactly equal to the (usual) sound mode computed in section \ssound.

As a final note, let us consider corrections to the dispersion relation \res.
Making the substitution $\omega\ra \omega=\pm\sqrt{{2\over 5}}q+\delta\omega$ and expanding eq. \eqqn\ to
first order in $\delta\omega$ yields the following solution
\eqn\deltaomega{\delta\omega={-1+2 \tilde{\gamma} +\ln{{q^2\over 10}}\over 10^{{3\over 2}}\tmu^2} q^3-
                  i {\pi\over 10^{{3\over 2}} \tmu^2}q^3+\OO(q^5)}
Notice that the imaginary part of the dispersion relation is of the type $-i q^3$ in contrast to
both the typical Fermi liquid behavior \PN\ and what was observed in \KarchFA, \ParnachevZS.


\newsec{Is there a sharp Fermi surface?}\seclab\sfs

In the previous section we observed a sound-like excitation
in the regime of vanishing temperature.
It is known that Fermi liquids possess such zero sound mode in
the collisionless regime.
It is associated with the deformation of the Fermi surface away
from the spherical shape.
The natural question is whether we can observe the existence of
the Fermi surface directly.
From the theory of normal Fermi liquids we learn that the jump
in the distribution function can indeed be observed as a singularity
in the retarded current-current Green's function.
More precisely, the two point function at $\omega=0$ behaves
like
\eqn\gsing{ G(\omega=0,q)\sim \left({q\over 2 q_F}-1\right) \log \left({q\over 2 q_F}-1\right)  }
as $q$ approaches the value of twice the Fermi momentum $q_F$.
In the previous section.
we focused on the massless excitation, but the equations can
be easily adapted to the case of $\omega=0$, $q$ finite.

In this section we analyze equation \eqfore\ for $\omega=0$.
It will be convenient to use the variable $x={\tilde\mu} y$, where
$\tilde\mu$ is defined in \tmudef\ and introduce the rescaled momentum $\tq$
via $\tq=q/{\tilde\mu}$.
Then eq.  \eqfore\ takes the form
\eqn\ewzero{  \p_x^2 E +{1\over x} {13 x^{10}-2\over 1+x^{10}} \p_x E-{\tq^2\over 1+x^{10}} E=0  }
In the near-horizon region, $x\gg1$ the solution of eq. \ewzero\
is a linear combination of the two solutions,
\eqn\ehorqz{  E^{(1)} \simeq 1,\qquad  E^{(2)}\simeq {1\over x^{12}}  }
We would like to argue that it is the second solution which is
physical.
Indeed, as the value of $q$ is taken to zero, fluctuation of the electric
field on the $D8$ brane corresponds to the infinitesimal change of the value of $\mu$.
Yet, in this static situation the boundary condition at the horizon
imposes $E(x\ra\infty)=0$.
In fact, in Appendix B we show that  $E(x\ra\infty)=0$ condition smoothly
connects to the incoming wave boundary conditions at the horizon, as $\omega$
is taken to be nonzero.

We can now use initial conditions at $x_{max}\ra\infty$ to integrate eq. \ewzero\
numerically all the way to the boundary.
Near the boundary $x=0$ the solution of \ewzero\ is given by
\eqn\elc{  E=\A F_I(x) + \B F_{II}(x)  }
where
\eqn\solewzero{ F_I(x)= \left(1 -{\tq^2 x^2\over2}-{\tq^4 x^4\over 8}\right)+\ldots,\quad
                F_{II}(x)=x^3+\ldots    }
\midinsert\bigskip{\vbox{{\epsfxsize=3in
        \nobreak
    \centerline{\epsfbox{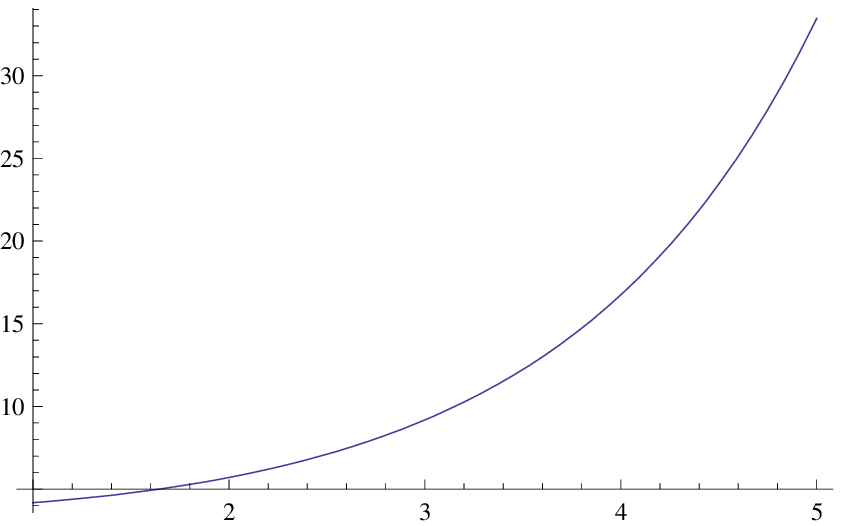}}
        \nobreak\bigskip
    {\raggedright\it \vbox{
{\bf Fig 1.}
{\it  Ratio ${\cal A}/{\cal B}$ [see eq. \elc] as a function of $\tq$.
}}}}}}
\bigskip\endinsert
\noindent
The two-point function is proportional to the ratio ${\cal A}/{\cal B}$, which can be
extracted from the numerical solution with the help of \solewzero.
The result is shown in Fig.1.
As evident from this figure, we do not observe any characteristic structure in the
wide region of $\tq$.
Note that this is the phase which exhibited such characteristic features of the Fermi
liquid as linear heat capacity and zero sound.
We discuss the significance of these observations in the next section.


\newsec{Discussion}

In this paper we  consider the \dfde\ system
in the limit $R_4\ra\infty$ which corresponds to a certain UV completion of
the NJL model at weak coupling.
Chemical potential for the fermions is turned on, and from the field theoretic
point of view formation of a Fermi liquid is a realistic possibility.
The holographic description is applicable at large coupling, and a phase
transition from the Fermi liquid to some other ground state may in principle happen
as coupling is varied.
However we observed a couple of features that suggest that such a phase transition
does not happen and we are dealing with a strongly coupled Fermi liquid-type system.

The most significant indication that this is the case is the linear
behavior of the density-dependent part of the heat capacity at
low temperatures \CvdlowT.
This is a very generic feature of the systems which have a step-like
distribution function at zero temperature.
At small temperature the number of excited states is proportional to $T$,
as well as the average energy of the excitation, leading to $E\sim T^2$
and $C_\upsilon\sim T$.
Note that this behavior is in sharp contrast with that of $D3-D7$ system
at small temperature, where $C_\upsilon\sim T^6$ \KarchFA.
This difference may be related to the existence of
charged bosons in the case of the  $D3-D7$ system.
Their condensation is presumably responsible for the different low temperature
behavior of $C_\upsilon$.
It is impressive that  the physics seems to vary smoothly between the
field theoretic and holographic regimes.

In section \sfluct\ we found a sound mode at vanishing temperature, the
zero sound. Again, such a feature is characteristic of Fermi liquids, where
zero sound is associated with the deformation of the Fermi surface away from
the spherical shape.
It is interesting that the speed of zero sound, given by \res\ is
equal to the speed of normal sound propagating  due to the
fermion matter in the hydrodynamic regime [see eq. \soundvelocity.]
Note that the speed of this ``brane sound'' differs from that of
the sound propagating on the stack of $D4$ branes by a factor of $\sqrt{2}$.
(The general formula for the speed of sound on the $Dp$ branes is
$\upsilon_s^2=(5-p)/(9-p)$, see e.g. \refs{\MasNG\CaceresTA-\MyersTB}.)
This is not a surprise since the fermionic degrees of freedom are localized
on a 3+1 dimensional defect.

In fact, as we show in appendix A, the value of the speed of zero sound
coincides with the one of first sound for the $D3-D7$ system as well.
There, the two expressions match as functions of the ratio
between the hypermultiplet mass and the chemical potential.
The fact that the speed of sound in these two regimes is the same
is in principle consistent with the Fermi liquid theory.
As reviewed in \ParnachevZS, when only the first two Landau parameters
$F_0$ and $F_1$ are turned on, the ratio of zero to first sound
velocities goes to one as $F_0$ and $F_1$ are taken to infinity.
For a generic large interaction $F(\vartheta)$, the ratio still goes
to a constant.
We leave the study of interpolation of the sound mode between the
collisionless and hydrodynamic regimes to future work.

It is interesting that the damping term in the dispersion relation
for the zero sound behaves like $-i q^3$.
This should be contrasted with the  $-i q^2$ damping observed in \KarchFA,\ParnachevZS.
Possible microscopic explanation of this behavior might be related to the
non-linear dispersion relation near the Fermi surface.
In fact\foot{We thank Edward Shuryak for pointing this out.}
this may also explain the absence of the apparent singularity in
the current-current two-point function at $\omega=0$, computed in section \sfs.
Indeed, the absence of the singularity, which one would generically
expect at $q=2 q_F$ is one of the main puzzles that we encounter.
It would be nice to have a phenomenological theory that would explain
the observations of the present paper.
It is clear that such a description must account for strong coupling,
since the relevant dimensionless parameter $\lambda \mu$ is necessarily
large for the DBI analysis to be applicable.
One obvious difficulty in applying Landau's theory of Fermi liquids is
the assumption that the number of particles is conserved as the strength
of the interaction is varied.
It is not obvious that this should be the case here.

Finally, we would like to make a comment about the dimensions of the D-branes.
In this paper we considered the \dfde\ system, which has problems with the
dilaton tadpole beyond the probe brane approximation for the $D8$ branes.
However we expect the physics to be similar for other D-brane systems,
for example  obtained from the \dfde\ by the T-duality.
It would be interesting to investigate these cases in more detail.

\bigskip
\bigskip

\noindent
{\bf Acknowledgements:} We thank A. Abanov, J. S. Caux, A. Karch, 
K. Schoutens, E. Shuryak, K. Skenderis, D. T. Son, T. Springer and  M.Taylor 
for very useful discussions and correspondence. M.K. is grateful to the Institute of Theoretical Physics at Stony Brook University where part
of this work was completed. A.P. is grateful to Harvard University, University of Chicago and MIT for hospitality
during the completion of this work. M.K. acknowledges support from NWO Spinoza Grant.

\bigskip
\bigskip

\appendix{A}{$D$3-$D$7 system and speed of sound.}\seclab\aa

In this appendix we compute\foot{After this Appendix had been written we received
\KimBV\ which contains similar result.} the speed of sound for the system studied in \ParnachevZS.
As already mentioned in section \srev, the speed of (normal) sound in liquids is defined as
\eqn\defzsths{  u^2=\left({\p P\over\p\epsilon} \right)_{\rho} }
with $P$ the pressure and $\epsilon$ the energy density. Evaluating the speed of sound is then
a trivial exercise once the grand canonical potential of the system is known.
Within the framework of gauge/gravity duality, the grand canonical potential
is identified with the DBI action evaluated on the specific configuration.
Here we are interested in the black hole embedding of the D3-D7 system at zero temperature.
Analytic results in this case, including evaluation of the action, are given in \KarchBR.
According to \KarchBR, but in the conventions of \ParnachevZS
\eqn\Xiths{S_{ren.}=-\NN \int_{0}^{\infty} \left({r^6\over \sqrt{r^6+d^2-c^2}}-r^3\right)={1\over 4}\NN  \gamma\left(\tmu^2-\tm^2\right)^2 }
where $\NN$, $\gamma$, $\tmu$ and $\tm$ are defined in section 3 of \ParnachevZS .

At zero temperature the pressure is equal and opposite to the grand canonical potential
$P=-\Xi$ while the energy density is equal to the free energy and given by
\eqn\energydts{\epsilon=\Xi+\mu\rho=-S_{D7}+\mu {\p S_{D7}\over\p \mu}={1\over 4}\NN\gamma (\tmu^2-\tm^2)(3\tmu^2+\tm^2)}
Given that pressure and energy density are functions of the chemical potential only\foot{Note that we take the hypermultiplet 
mass $m$ as a fixed parameter of our ensemble.}  
we can rewrite \defzsths\ as
\eqn\zsthstwo{u^2=\left({\p P\over\p\epsilon} \right)=\left({\p P\over\p\mu}\right)\left({\p \mu\over\p\epsilon} \right)}
and using \Xiths\ as well as \energydts\ we conclude that
\eqn\zsths{u^2={\tmu^2-\tm^2\over 3\tmu^2-\tm^2}}
Observe that the speed of zero sound computed in \ParnachevZS\ is exactly equal to \zsths.

\appendix{B}{The boundary condition at the horizon in the limit $\omega\ra 0$.}\seclab\ab

In section \sfs, we numerically computed the retarded Green's function for the density operator
and investigated its behavior for different values of $q$. Given that the singularity which signals the
presence of a Fermi surface sits at $\omega=0$, the calculation was performed by setting $\omega=0$
in \eqrorlong. However, the appropriate boundary conditions at the horizon when $\omega=0$ are not known.
Here, we will show that in the limit $\omega\ra 0$ the incoming wave boundary condition reduces to
\eqn\bcwzero{\lim_{y\ra \infty}E(y)=0}

Let us investigate the region close to the horizon henceforth defined as the regime where $\tmu y\gg 1$.
Given that we want to study the behavior of the solution at very small frequencies, we will not require
${\omega^2\over q^2} \tmu^{10}y^{10}$ to be large. Eq. \eqrorlong\ in this case reduces to
\eqn\eqhorwq{\ddot{E}+\left({3\over y}+{10\over y} {1\over 1 -{\omega^2\over q^2} \tmu^{10}y^{10}}\right)\dot{E}+
                  \omega^2 \left(1-{1\over {\omega^2\over q^2}\tmu^{10}y^{10}}\right)E=0    }

Observe that when ${\omega^2\over q^2} \tmu^{10}y^{10}\gg 1$, \eqhorwq\ further reduces to \eqh.
In this case, the incoming wave boundary condition singles out one of the solutions, given by
\eqhsol. If in addition $\omega y\ll 1$, \eqhsol\ behaves like
\eqn\eqhsolsmallw{E(y)\simeq {A\over y^2}+A \omega^2 \left[{1\over 4}\left(1-2\tilde{\gamma}+i \pi+2 \ln{2}\right)-{1\over 2}\ln{\omega y}\right]}

On the other hand, when ${\omega^2\over q^2} \tmu^{10}y^{10}\ll 1$, eq. \eqhorwq\ becomes
\eqn\eqhorwzero{\ddot{E}+{13\over y}\dot{E}=0}
This is precisely \eqrorlong\ in the vicinity of the horizon for $\omega=0$.
Its general solution is expressed as
\eqn\solhorwzero{E(y)=-{D\over y^{12}}+F}
Clearly, one of the solutions vanishes for large $y$ while the other behaves as a constant.
Which of the two corresponds to \eqhsolsmallw ?

It is not difficult to see, that there exists an additional, intermediate region where
\eqhorwq\ is analytically tractable. Consider for instance the behavior of the last term
for both large and small $y$.
\eqn\lastterrm{\omega^2 \left(1-{1\over {\omega^2\over q^2}\tmu^{10}y^{10}}\right)\simeq
\left \{
\eqalign{
       &\quad {\omega^2} \qquad\qquad {\omega^2\over q^2}\tmu^{10}y^{10} \gg 1 \cr
       &-{q^2 \over \tmu^{10}y^{10}} \qquad {\omega^2\over q^2}\tmu^{10}y^{10} \ll 1}
         \right. }
It is obvious that this term is negligible so long as\foot{Here we tacitly assumed that $\left({q\over\tmu}\right)>1$.
If this is not the case, eq. (A.8) extends its regime of validity in the region ${\omega^2\over q^2} \tmu^{10}y^{10}\ll 1$ as well.
Repeating the same analysis we can easily see that the final result remains unaltered.}
\eqn\rtwocondition{\left({q\over\tmu}\right)^{{1\over 4}} {1\over\tmu} \ll y\ll {1\over\omega}}
Consequently, eq. \eqhorwq\ reduces to
\eqn\eqhornolastterm{\ddot{E}+\left({3\over y}+{10\over y} {1\over 1 -{\omega^2\over q^2} \tmu^{10}y^{10}}\right)\dot{E}=0}
with general solution
\eqn\solhornolastterm{E(y)={B\over y^2}\left(1- {1\over 6}{q^2\over \omega^2}{1\over\tmu^{10} y^{10}}\right) +C}
For large $y$ we can neglect the term which behaves like $y^{-12}$ to get
\eqn\solhornltlargey{E(y)\simeq {B\over y^2}+C }
Similarly, in the region of small $y$ \solhornolastterm\ becomes
\eqn\solhornltsmally{E(y)\simeq -B{1\over 6}{q^2\over \omega^2}{1\over\tmu^{10} y^{12}}+C}

In principle, we should proceed to match eq. \solhornltlargey\ to \eqhsolsmallw\ and eq. \solhornltsmally\ to \solhorwzero.
Observe however, that \solhornltlargey\ does not contain the logarithmic term present in \eqhsolsmallw.
To recover the logarithmic behavior we need to solve \eqhornolastterm\ in the near horizon region to
next order in $\omega$. We should therefore consider the following equation
\eqn\difcor{\ddot{E}+{3\over y}\dot{E}=\omega^2 \left({B\over y^2}+C\right)}
Solving \difcor\ yields
\eqn\solcor{E(y)={B\over y^2}+C-{1\over 2} B \omega^2 \ln{y}+\OO(y^2)}
and combining all the above we deduce that
\eqn\FoverD{{F\over D}=6 \omega^4 {\tmu^{10}\over q^2} {1\over 4}\left(1-2\tilde{\gamma}+i \pi-2 \ln{{\omega\over 2}}\right)}

It is clear, that \FoverD\ vanishes when $\omega=0$ and the other parameters are kept fixed.
Note that taking the limit $\omega\ra 0$ is completely justified given that the matching technique
is valid in the regime
\eqn\mtvl{\omega\ll \tmu \left({\tmu\over q}\right)^{{1\over 4}}}

\footatend\vfill\supereject\immediate\closeout\rfile\writestoppt
\baselineskip=14pt\centerline{{\bf References}}\bigskip{\frenchspacing%
\parindent=20pt\escapechar=` \input refs.tmp\vfill\eject}\nonfrenchspacing
\end